\documentclass[twocolumn]{aastex63}
\usepackage{amsmath}
\usepackage{verbatim}
\usepackage{graphicx,url}
\usepackage{amsmath,amssymb}
\usepackage{hyperref}
\hypersetup{colorlinks,
linkcolor=blue, 
urlcolor=magenta, 
anchorcolor=blue,
citecolor=blue
}
\usepackage{subfigure}
\usepackage[normalem]{ulem} 
\usepackage{soul} 

\definecolor{greenW}{rgb}{0.0, 0.55, 0.1}

\begin{document}

\title{Recovering Cosmic Microwave Background Polarization Signals with Machine Learning}

\author{Ye-Peng Yan}
\affiliation{Department of Astronomy, Beijing Normal University, Beijing 100875, China; xiajq@bnu.edu.cn}

\author{Guo-Jian Wang}
\affiliation{School of Chemistry and Physics, University of KwaZulu-Natal, Westville Campus, Private Bag X54001,Durban, 4000, South Africa}
\affiliation{NAOC-UKZN Computational Astrophysics Centre (NUCAC), University of KwaZulu-Natal, Durban, 4000, South Africa}

\author{Si-Yu Li}
\affiliation{Key Laboratory of Particle Astrophysics, Institute of High Energy Physics, Chinese Academy of Science, P. O. Box 918-3, Beijing 100049, China}

\author{Jun-Qing Xia}
\affiliation{Department of Astronomy, Beijing Normal University, Beijing 100875, China; xiajq@bnu.edu.cn}

\begin{abstract}
Primordial B-mode detection is one of the main goals of current and future cosmic microwave background (CMB) experiments. However, the weak B-mode signal is overshadowed by several Galactic polarized emissions, such as thermal dust emission and synchrotron radiation. Subtracting foreground components from CMB observations is one of the key challenges in searching for the primordial B-mode signal. Here, we construct a deep convolutional neural network (CNN) model, called \texttt{CMBFSCNN} (Cosmic Microwave Background Foreground Subtraction with CNN), which can cleanly remove various foreground components from simulated CMB observational maps at the sensitivity of the CMB-S4 experiment. Noisy CMB Q (or U) maps are recovered with a mean absolute difference of $0.018 \pm 0.023\ \mu$K (or $0.021 \pm 0.028\ \mu$K). 
To remove the residual instrumental noise from the foreground-cleaned map, inspired by the needlet internal linear combination method, we divide the whole data set into two ``half-split maps,'' which share the same sky signal, but have uncorrelated noise, and perform a cross-correlation technique to reduce the instrumental noise effects at the power spectrum level. We find that the CMB EE and BB power spectra can be precisely recovered with significantly reduced noise effects. Finally, we apply this pipeline to current Planck observations. As expected, various foregrounds are cleanly removed from the Planck observational maps, with the recovered EE and BB power spectra being in good agreement with the official Planck results.
\end{abstract}

\section{Introduction}\label{sec:Introduction}

The next frontier in cosmic microwave background (CMB) experiments is to precisely measure the anisotropy of its polarization state. Since the primordial CMB B-mode signal is a clear sign of the primordial gravitational waves of quantum origin predicted by inflation \citep{Kamio:2016}, several next-generation CMB experiments have been proposed or are under construction—such as the CMB-S4 project \citep{Abazajian:2019}, LiteBIRD satellite \citep{Hazumi:2019, Suzuki:2018}, the Simons Observatory \citep{Ade2019} and AliCPT \citep{Li2017} —which are designed to have multifrequency coverage and very high sensitivity, to search for the B-mode signal. However, this signal is much weaker for all frequency bands than the Galactic polarized radiation on degree and larger angular scales, \citep{Krachmalnicoff:2016,Krachmalnicoff:2018}, so component separation is one of the critical challenges for CMB data analysis.

Typical methods for component separation can be broadly divided into two categories: ``parametric" methods, such as \texttt{Commander} \citep{Eriksen:2008} and \texttt{XFORECAST} \citep{Errard:2016,Stompor:2016}, and ``blind" methods, such as the Internal Linear Combination (ILC) \citep{Tegmark:2004,Kim:2009,Sudevan:2017}. The parametric methods are highly dependent on foreground modeling. Slightly inaccurate foreground modeling could lead to strong bias in the reconstruction of the CMB B-mode signal, due to the huge gap between the amplitude of the Galactic foreground contamination and the primordial B-mode signal \citep{Armitage-Caplan:2012,Remazeilles:2016,Hensley:2018}. On the other hand, the blind methods mainly exploit the minimal prior information for the blackbody spectrum of CMB radiation. By discarding existing models, blind methods are able to deal with unknown and complex foreground contamination, to reconstruct clean CMB maps.

\begin{figure*}
\begin{center}
	\includegraphics[width=0.9\hsize]{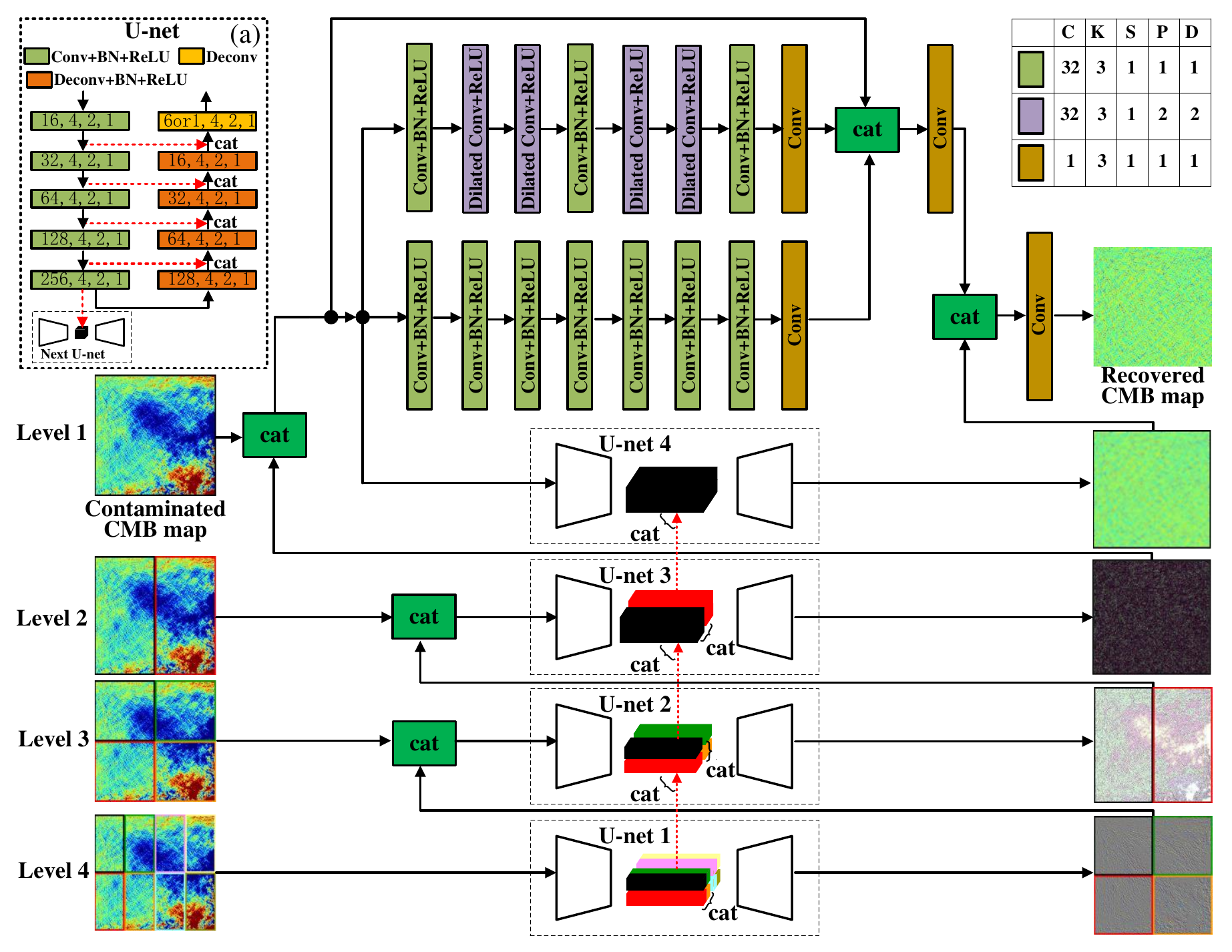}
\end{center}
\vspace{-0.4cm}
\caption{Network architecture. "Conv" stands for convolution; "Dilated Conv" stands for dilated convolution; "BN" stands for batch normalization; and "ReLU" stands for leaky rectified linear unit (negative slope set to 0.4), respectively, while "cat" denotes concatenation operation. "C," "K," "S," "P," and "D" represent the number of output channels, the kernel (filter) sizes, the stride, the padding, and the dilation of the convolutional layer, respectively. Panel (a) is the U-net architecture, where the numbers from left to right are the values of the output channel, the kernel size, the stride, and the padding, respectively. "Deconv" stands for the deconvolutional layer. The input images consist of observational (Q or U) maps at multiple frequencies concatenated along the channel dimension. The input images pass through the convolution layers of each level, before the output image (a foreground-cleaned Q or U map) is finally obtained.}
\label{figure_net}
\end{figure*}

With the remarkable progress of computer science over recent years, machine learning has shown excellent capabilities in the image processing field—in terms of image recognition, denoising, and deblurring—which have increasingly been applied to astrophysical research \citep{Mehta:2019,Fluke:2020,Wang:2020}.  In particular, machine learning has been used to extract CMB temperature maps from observational data \citep{Petroff:2020,Wang:2022,Casas:2022}, recognize foreground models \citep{Farsian:2020}, and extend CMB foreground models \citep{Krachmalnicoff:2021}. Here, we present in this paper a novel method for polarized foreground removal, based on a machine-learning technique. The basic idea is to optimize the parameters in a neural network by training it with pregenerated CMB data sets. Our neural network aims to find the mapping between multifrequency microwave observations and clean CMB polarization signals.

\section{Methodology}\label{sec:Methodology}
\subsection{Network Architecture}
For image processing tasks, the convolutional neural network (CNN) is the most ubiquitous type of feed-forward neural network, its core being the convolutional layer \citep{Dumoulin:2016,Mehta:2019}. Each convolutional layer accepts the feature images from the previous layer as inputs, then convolves them with many local spatial filters (or kernels), whose values are parameters to be learned, before finally providing outputs to the next layer, following a nonlinear activation function. The output size of a convolutional layer is controlled by three hyperparameters: the number of output channels, the stride, and the amount of zero padding. The stride is defined as the distance in pixels between the centers of adjacent filters. In addition to standard convolution, dilated convolution can capture more contextual information, by enlarging the receptive field size of the filter \citep{Yu:2015}. One can design a network architecture by connecting many convolutional layers. The network architecture therefore consists of a stack of nonlinear parameters, which can convert complex problems into parameter optimization, by minimizing the loss function when training the network on a data set. A successful architecture that is widely used for image processing tasks is U-net \citep{Ronneberger:2015}, which encodes relevant information from the input maps onto smaller maps, decodes that information to form the output maps, then adds extra shortcuts (skip connections) between the encoding and decoding layers, to allow for the propagation of small-scale information that might be lost when the sizes of the images decrease.

The network model used in this work is a multi-patch hierarchy network \citep{Nah2016,Waqas2021}, based on CNN, which consists of four levels, as shown in Figure 1. Contaminated CMB maps at multiple frequencies are fed to the network model, going through each level and outputting a foreground-cleaned CMB map at the selected frequency band. Each level contains a U-net (panel (a)), with five convolutional layers (called the encoder) and five deconvolutional layers (called the decoder), and there are connections between the levels. For levels 4, 3, and 2, the input maps are sliced into eight, four, and two overlapping patches of equal size, respectively, and the patches are fed to the corresponding U-net, to extract information about the CMB on different scales. Two adjacent feature maps in the encoder will be spatially concatenated after the activation function (the leaky ReLU function), then the new feature maps will be passed to the mirrored deconvolutional layer (or the upper level), to be concatenated along the channel dimension, via skip connections (the red dashed lines). The output of the decoder is also passed to the upper level, to be concatenated along the channel dimension. There are three branches in level 1: one is the U-net, and the other two are based on the BRDNet network \citep{Tian:2020} which consists of dilated and standard convolutions. The contaminated CMB maps and feature maps from the lower levels are concatenated (along the channel dimension), so that they can be passed to these three branches for information to be extracted, before the foreground-cleaned CMB map will be finally obtained.

The network will be optimized by minimizing a loss function, which is defined as
\begin{align}
	\mathcal{L} = \mathcal{L}_{\rm MAE} + \beta \mathcal{L}_{\rm FFT},
\end{align}
where $\mathcal{L}_{\rm MAE}$ is the least absolute deviation (MAE, also called L1 loss), $\mathcal{L}_{\rm FFT}$ is a physics-guided loss based on the fast Fourier transform (FFT), and $\beta$ is a coefficient representing the contribution of $\mathcal{L}_{\rm FFT}$ to the total loss and we set $\beta=1$ throughout the paper. Specifically, for the training set with $S$ pairs of samples $\{x_i, y_i\}_{i=1}^S$ ($x$ is the contaminated image, and $y$ is the corresponding ground truth), the L1 loss is
\begin{align}
	\mathcal{L}_{\rm MAE}&=\frac{1}{N}\sum_{n=1}^{N}\left[ \frac{1}{WH}\sum_{w=1}^{W}\sum_{h=1}^{H}(|I^{n}_{w,h}-y^{n}_{w,h}|)\right],
\end{align} 
where $N$ is the batch size, $H$ ($W$) is the height (width) of the images in pixels, and $I = f(x)$ ($f(\cdot)$ is the network model) is the predicted image. The formula of $\mathcal{L}_{\rm FFT}$ is
\begin{align}
	\mathcal{L}_{\rm FFT}&=\frac{1}{N}\sum_{n=1}^{N}\left[ \frac{1}{WH}\sum_{w=1}^{W}\sum_{h=1}^{H}(|A_{\rm F}(I^{n}_{w,h})-A_{\rm F}(y^{n}_{w,h})|)\right],
\end{align}
where $A_{\rm F}$ is the amplitude of FFT, which has the form of 
\begin{align}
	A_{\rm F}(I)& = \sqrt{Re[{\rm FFT}(I)]^2+Im[{\rm FFT}(I)]^2},
\end{align}
where $Re[\cdot]$ and $Im[\cdot]$ denote the real and imaginary parts, respectively.

\subsection{Training data}
\label{training_data}
The CNN method is a supervised machine-learning algorithm that requires the use of a training data set with known truth values. We simulate the training data sets using the Python Sky Model (\texttt{PySM})\footnote{\url{https://github.com/bthorne93/PySM_public}} \citep{Thorne:2017}. In particular, the sky emissions that are considered in our simulation consist of synchrotron emission, thermal dust emission, polarized anomalous microwave emission (AME), and the CMB. For the CMB simulation, we first use the \texttt{CAMB}\footnote{\url{https://github.com/cmbant/CAMB}} package to obtain the CMB power spectra, then we use the Synfast function of the public software \texttt{Healpy}\footnote{\url{https://github.com/healpy/healpy}} \citep{Zonca2019} (a python wrapper
of HEALPix\footnote{\url{https://healpix.sourceforge.io/downloads.php}} \citep{Gorski2005})  to generate the CMB map realizations, with $N_{\rm side}=512$. We consider a standard $\Lambda$ cold dark matter model, and take the best-fit values and standard deviations from the Planck 2015 data \citep{PlanckCollaboration:2016}.  Here, we do not consider the CMB lensing effect during the map making, but we check that the overall conclusions stand when considering the CMB lensing. The foreground simulation is based on the parametric models listed below:
\begin{equation}
	\begin{split}
		T_s(\hat{n},\nu)&=A_{s_{\nu_{0}}}(\hat{n})\left(\frac{\nu}{\nu_{0}}\right)^{\beta_s(\hat{n})}, \\
		T_d(\hat{n},\nu)&=A_{d_{\nu_{0}}}(\hat{n})\left(\frac{\nu}{\nu_{0}}\right)^{\beta_d(\hat{n})}B(\nu,T_d(\hat{n})),\\
		I_a(\hat{n},\nu)&=A_{\nu_{0,1}}(\hat{n})\epsilon(\nu,\nu_{0,1},\nu_{p,1}(\hat{n}),\nu_{p_0})\\
		&+A_{\nu_{0,2}}(\hat{n})\epsilon(\nu,\nu_{0,2},\nu_{p,2}(\hat{n}),\nu_{p_0}),\\
		Q_a(\hat{n},\nu)&=fI_{a}{\rm cos}(2\gamma_{353}),\\
		U_a(\hat{n},\nu)&=fI_{a}{\rm sin}(2\gamma_{353}),
	\end{split}
\end{equation}
where $T_s$ and $T_d$ represent the synchrotron and thermal dust emission temperature as functions of orientation $\hat{n}$ and frequency $\nu$. $I_a$, $Q_a$, and $U_a$ are the intensity, Q and U polarization of AME. $A$s represent various amplitudes at the pivot frequencies $\nu_0$s and $\beta$s are the spectral index. $B$ represents the black body emission law for thermal dust emission.  All foreground components are simulated using \texttt{PySM}, with the same resolution as the CMB simulations. Specifically, we adopt the s1 model for the synchrotron emission, the d1 model for the thermal dust emission, and the s2 model for the AME in \texttt{PySM}.

\begin{table}
\centering
\renewcommand\tabcolsep{15pt}
\caption{Frequencies and instrumental specifications of  the CMB-S4 experiment \citep{Abazajian:2019}.\label{table_1}}
\begin{tabular}{c|c|c}
\hline \hline
Frequency  & Sensitivity & FWHM  \\
(GHz) & ($\mu$K-arcmin) & (arcmin)\\
\hline
85.0 & 1.31 & 25.5 \\
95.0 & 1.15 & 22.7 \\
145.0 & 1.78 & 25.5 \\
155.0 & 1.91 & 22.7 \\
220.0 & 4.66 & 13.0 \\
270.0 & 7.99 & 13.0 \\
\hline \hline
\end{tabular}
\end{table}

\begin{figure*}
	\includegraphics[width=1.0\hsize]{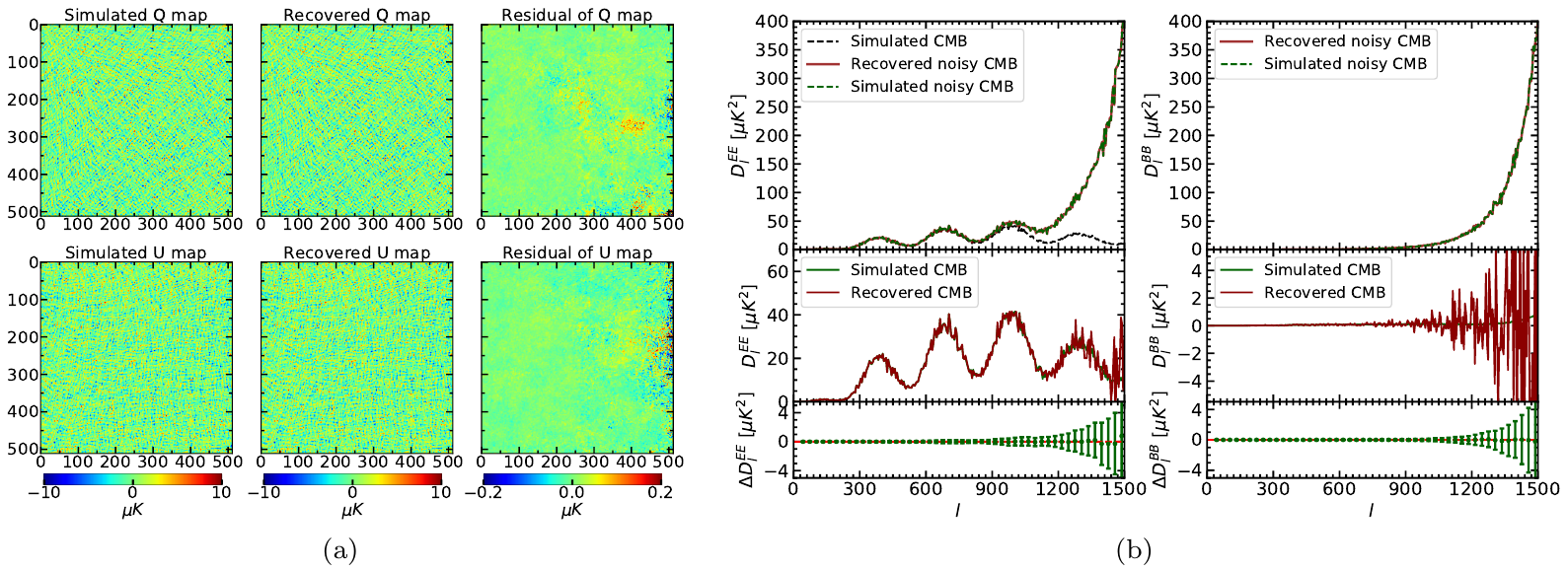}
	\vspace{-0.4cm}
	\caption{ (a) The recovered noisy CMB polarization maps for the partial-sky observation of the CMB-S4 experiment. The simulated maps are the pure CMB signal with instrumental noise at 220 GHz. The recovered maps are the noisy CMB maps recovered by {\tt CMBFSCNN}. The corresponding residual maps are also represented. (b) The two upper panels show the EE and BB power spectra of the recovered noisy CMB maps for the CMB-S4 experiment. The two lower panels show the final EE and BB power spectra, after the denoising step. We also show the deviations of the recovered power spectra from the input fiducial ones.}
	\label{figure2}
\end{figure*}

In machine learning, the capability of a trained neural network to forecast an unseen data set is called generalization. We always hope that the generalization will be strong enough to handle the real data, which means that one needs to generate training data that are close enough to the real world. However, the foreground emission is too complex \citep{Finkbeiner:1999,Kogut:2007,Kogut:2012} to be parameterized, and this is reflected as the complicated spatial variation of the amplitude and spectral index in the power-law model of synchrotron, the modified blackbody model of thermal dust, and the model of AME. In practice, while generating the training data, we manually add uncertainties to the parameters, to enhance the generalization of our network. Specifically, for the CMB realizations, the values of the cosmological parameters are treated as independent Gaussian random variables, where the mean values and standard deviations are taken from the best-fit values and standard deviations of the Planck 2015 results. For the foreground realizations, we multiply the value of each pixel in the amplitude template map $A$ (and spectral index map $\beta$) by a random number with an average value of 1 and a standard deviation of 0.1 (0.05 for a spectral index).

The inputs to the network model are the beam-convolved observational maps at multiple frequencies, which contain the CMB, foregrounds, and instrumental noise, while the outputs are the beam-convolved CMB maps plus noise maps, meaning that our network is designed to remove the foregrounds only. As the neural network requires 2D data for the inputs and outputs, we first divide each input map into 12 patches, in the NESTED ordering scheme, and directly fill each data patch with an $N_{\mathrm{side}}\times N_{\mathrm{side}}$ square grid, then combine these grids into a 2D map. To obtain the output maps in HEALPix format, one can perform an inverse process on the network's outputs. More relevant information can be found in \citet{Wang:2022}. Finally, the angular power spectra $C_\ell$ of the partial-sky maps are calculated using \texttt{NaMaster}\footnote{\url{https://github.com/LSSTDESC/NaMaster}} \citep{Alonso2019}.

%
\section{Application to CMB experiments}\label{sec:Application}
\subsection{Removing foregrounds on CMB-S4}
Firstly, we test our method on the simulated data with the performance of CMB-S4 experiment.  We simulate 1000 beam convolved emission maps, and 300 white noise maps for the training set at six frequency bands with NSIDE of 512.  Here, we assume that the instrument noise is Gaussian and white. For each frequency band, these 300 white noise maps are randomly added to 1000 beam convolved emission maps. In total, the training set consists of 1000 observed maps at six frequency bands. The generation of simulated maps is described in section \ref{training_data}, while the corresponding frequency band and instrumental properties \footnote{\url{https://cmb-s4.uchicago.edu/wiki/index.php/Survey_Performance_Expectations}} used here are summarized in Table \ref{table_1}. For the test set, 300 sky emission sets and 300 noise sets are generated, similar to the training sets, but with different random seeds and parameter values. Note that we set the network output as a beam-convolved CMB map with a noise map at 220 GHz. We adopt Adam \citep{Kingma:2014} as the optimizer and set the learning rate to 0.01 initially which gradually decreases to $10^{-6}$ over the iterations. We perform 20,000 iterations in the neural network, by minimizing the loss function. We train our network model on two NVIDIA Quadro GV100 GPUs, with a batch size of 14, and it takes $\sim 14$ hr for one network model.

We choose a patch with one-twelfth the size of sky and $512\times512$  pixels to train the network, then apply the trained neural network to a sample on the testing set. The results are shown in panel (a) of Figure \ref{figure2}. At first glance, the recovered polarization Q/U maps are quite similar to the input simulated maps, qualitatively. In order to evaluate the performance of our network quantitatively, we use the mean absolute difference (MAD), with the following general formula:
\begin{equation}
	\begin{split}
		\sigma_{\rm MAD} &= \frac{1}{N}\sum_{i}^{N}\left|X_i - Y_i\right|,
	\end{split}
\end{equation}

where $N$ is the number of pixels, $X$ and $Y$ represent the predicted and ground-truth maps. The MAD between the recovered CMB Q/U maps and the target Q/U maps (true CMB maps plus noise maps) is calculated, and we can obtain recovered Q and U maps with the MAD values: $0.0187\ \mu$K and $0.020\ \mu$K, respectively.  We also calculate the average MAD values for 300 testing sets: $0.018 \pm 0.023\ \mu$K for Q map and $0.021 \pm 0.028\ \mu$K for U map, respectively,  meaning that our network model has good generalization. We also calculate the average values of MAD for the training sets: $0.016 \pm 0.021\ \mu$K for Q map and $0.021 \pm 0.027\ \mu$K for U map, respectively,  consistent with the results of the test set. Note that the relative contributions of the noise and foreground in the calculations of the MAD values are not clear, because the noise is included in the calculation of the MAD values. We also calculate the MAD values between the recovered CMB Q/U maps minus the noise maps and the target Q/U maps minus the noise maps. We find that these MAD values are almost identical to the MAD values without the subtraction of the noise maps, implying that the MAD values could be used to estimate the goodness of removing the foregrounds. Furthermore, from the perspective of power spectra (EE and BB) recovery, as shown in panel (b) of Figure \ref{figure2}, the CMB plus noise signal can be accurately recovered, meaning that the foregrounds are the main contributors to the MAD values. The final target of the CMB pipeline is accurately recovering the CMB signal alone. These MAD values are about two orders of magnitude smaller than the true CMB Q/U maps, on average, implying that the deviations between the recovered maps and the target ones are very small.

The performance of the recovered maps on the right is obviously worse than that for other areas, as shown in panel (a) of Figure \ref{figure2}, since the region on the right is very close to the galactic plane and suffers from foreground contamination badly. It is very difficult to perfectly remove the total foreground from these heavily contaminated areas.

Besides the comparison at the map level, we also compare the EE and BB power spectra of the output recovered maps and the input fiducial maps, as shown in the upper two panels of panel (b) of Figure \ref{figure2}. The dotted black lines represent the theoretical power spectra with pure CMB signals, the green lines show the target power spectra, including the CMB signal and instrumental noise, and the red lines show the power spectra that are obtained from the output recovered maps. Apparently, the recovered maps have almost identical power spectra to the targets, meaning that our CNN network can cleanly remove the foreground contamination at both the map level and the power spectrum level. Note that the EE and BB spectra increase sharply at scales of  $\ell>1000$, because the power spectra include noise and are divided by the Gaussian beam window function, in order to remove the beam effects.

\begin{figure*}
	\includegraphics[width=1.0\hsize]{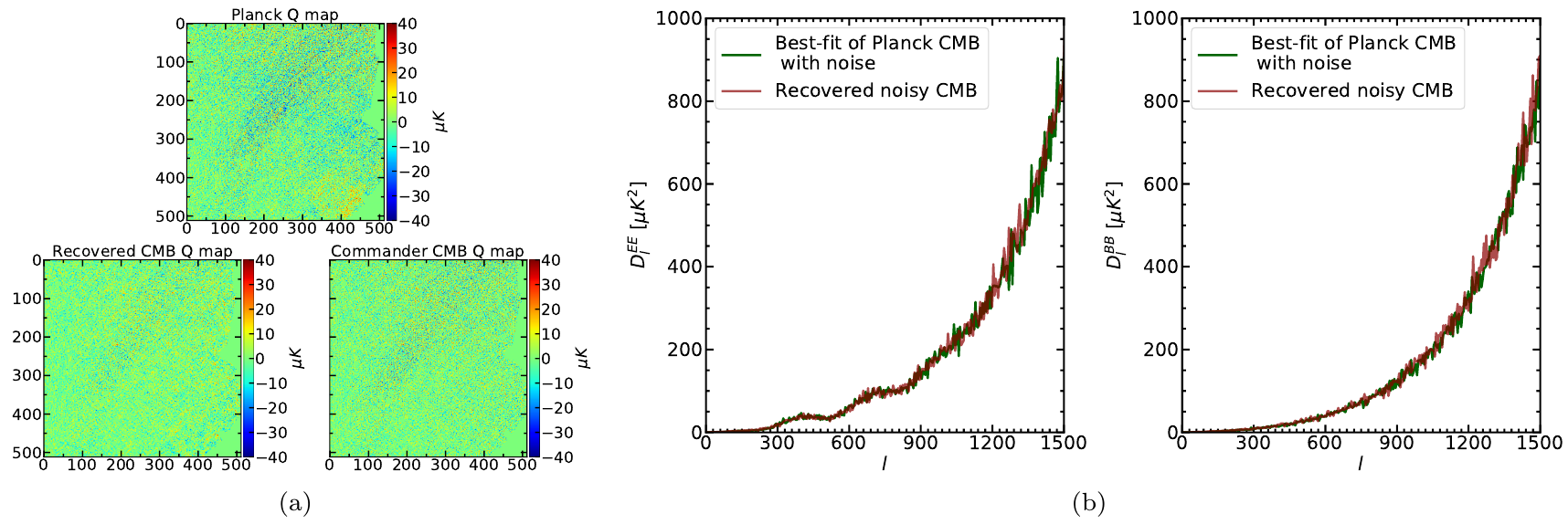}
	\vspace{-0.4cm}
	\caption{(a) The Planck Q map is the even-ring Q map for the Planck observation at 143 GHz. The recovered CMB Q map and the Commander CMB Q map are the network's output Q map and the Commander Q map. (b) Angular power spectra of the recovered Planck CMB maps.}
	\label{figure3}
\end{figure*}

\subsection{Denoising for Recovered Maps}
In our method, the output CMB maps obtained from the neural network still contain the noise effects from the instrument. \citet{Wang:2022} found that it is still very challenging to directly remove the noise effects at map level using CNN methods. For this network, then, we leave the noise in the output recovered maps and try to suppress the noise effects at the power spectrum level. Inspired by \citet{Krachmalnicoff2022}, we divide the whole dataset into two ``half-split (HS) maps" which share the same signal, but have uncorrelated noise, and calculate the cross-correlation power spectra of these two HS maps. Note that the noise of the HS maps is enhanced by a factor of $\sqrt{2}$, relative to the sensitivity listed in Table \ref{table_1}. Due to the uncorrelated noise, the noise effects almost disappear in the cross-correlation power spectra, but the CMB signal remains.

We plot the obtained cross-correlation power spectra in the lower two panels of panel (b) of Figure \ref{figure2}. The red lines represent the recovered EE and BB power spectra obtained from cross-correlating the two HS maps, which are in good agreement with the fiducial CMB signal (green lines). The noise effects are now significantly suppressed at the level of the power spectra. We also plot the difference, $\Delta D_{\ell,{\rm CNN}} =D_{\ell,{\rm recover}} - D_{\ell, {\rm true}}$, between the recovered and fiducial power spectra at the bottoms of the two panels, where the error bars represent the mean standard deviations over all the testing sets. We can see that the difference  $\Delta D_{\ell,{\rm CNN}}$ is quite consistent with zero, and as the multipole $\ell$ gets larger, the error bars are increasing, due to beam effects.

In addition, we also use  $R^2 = 1-\sigma_{\rm CNN}^2/\sigma^2$ to evaluate the performance of denoising on the power spectrum. Here, $\sigma_{\rm CNN}^2$ is calculated as
\begin{equation}
	\begin{split}\label{MS}
		\sigma_{\rm CNN}^2 &= \frac{1}{N}\sum_{i}^{N}\left(X_i - Y_i\right)^2,
	\end{split}
\end{equation}
where $N$, $X$ and $Y$ are the maximum multipoles ($\ell_{\rm max}=1500$), $D_{\ell,{\rm recover}}$ and $D_{\ell, {\rm true}}$, respectively. $\sigma^2$ is the variance of the target power spectrum, which is the one from the input fiducial map here. $R^2=1$ means that the recovered power spectrum is exactly the same as the target power spectrum. The closer $R^2$ is to 0, the worse the fitting performance is. In practice, we obtain the $R^2$ for the recovered EE power spectrum after denoising: $R^2= 0.98\pm 0.01$ (68\% C.L.) at all scales and $R^2= 0.9996\pm 0.0001$ (68\% C.L.) at $\ell<1200$, which shows that the recovered spectrum is in good agreement with the fiducial one. For the BB power spectrum, it is worthy noting that the uncertainties will obviously increase at $\ell>800$, which will significantly affect the calculation of $R^2$. Therefore, if we only consider the BB power spectrum at scales $\ell<800$, we can obtain the $R^2$: $R^2=0.89\pm0.095$, which implies we can obtain accurate measurement on the CMB B-mode signal at scales $\ell<800$ for the CMB-S4 experiment.


\subsection{Test with Planck CMB Maps}
To further test the capabilities of our network, we apply this pipeline to current Planck observations. We generate 1000 training sets of simulated observation maps for four High Frequency Instrument (HFI) channels using \texttt{PySM}, including 100, 143, 217, and 353 GHz. Note that these maps have already been convolved by the effective Gaussian beam for each band: $9'.66$, $7'.27$, $5'.01$, and $4'.86$, respectively. In addition, we take 300 sets of noise and systematic residual maps from the Planck Legacy Archive (PLA)\footnote{\url{http://pla.esac.esa.int/pla}}.
Similar to our process for CMB-S4, we randomly add the noise and systematic maps to the signal maps, to generate the input CMB observational maps. We also apply the Planck mask \texttt{COM\_Mask\_CMB-common-Mask-Pol\_2048\_R3.00.fits} from the PLA to the simulated maps, to eliminate the heavily contaminated area in the galactic plane.

Similar to the HS maps that were used before for the denoising, we here choose the "half-ring" (HR) maps provided by the Planck collaboration. The first and second halves of the stable pointing data share the same sky signal, since they come from the same scanning pattern, but have uncorrelated noise. In the Planck data processing, the HR maps have an important role in the noise estimation \citep{Zacchei:2011,PlanckCollaboration:2014, PlanckCollaboration:2014b} and component separation process \citep{PlanckCollaboration:2014c}. We will cross-correlate the Planck HR maps to derive the noise-removed power spectra.

Given that maps of multiple bands will be used during the foreground removal, this means that the noise of the output sky map should be less than those of each single frequency map, thanks to the accumulation of information. In order to generate the noise components of the target maps for the training sets, we directly compute the target noise map through the polarization ILC method \citep{Tegmark:2004,Kim:2009,Fernandez-Cobos:2016iud,Zhang:2021odl}, where the processed map can be written as:
\begin{equation}
	\hat{Q}(p)\pm i\hat{U}(p) = \sum_{f} \left(\omega^R_f\pm i\omega^I_f\right)\left(Q_f(p)\pm i U_f(p)\right)~.
\end{equation}
As an un-biased estimator, corresponding linear weights should satisfy the following conditions, reads:
\begin{align}
	\sum_f\omega^R_f=1, \sum_f\omega^I_f=0~,
\end{align}
and can be obtained by minimizing the variance of ${|\hat{Q}+i\hat{U}|}^2$. $p$ and $f$ stand for the pixel index and the frequency channel. We perform the polarization ILC on each training set—consisting of the CMB, the foreground emission from the \texttt{PySM} package, noise and systematic from PLA simulations, to get the corresponding weight coefficients $\omega^R$ and $\omega^I$. We then obtain the target noise map as the weighted sum of the noise maps for the four channels.

We set up the neural network with the input maps that we use in the polarization ILC procedure, with the target output CMB map being convolved by the beam at 143 GHz, together with the ILC noise map from the four noise maps for the HFI channels. Note that we apply the Planck mask on both the input and target output maps, to eliminate the heavily contaminated area in the galactic plane.

The training process is also similar to the process for CMB-S4, and we choose the same sky patch, with one-twelfth the size of sky and $512\times512$ pixels, to train the network. Once the network is trained, the real frequency maps of the even and odd rings from the PLA are fed into the trained network. After using the CNN network, we obtain a recovered noisy Planck Q map, as shown in panel (a) of Figure \ref{figure3}. When comparing with the input Planck Q map, we can see that a lot of significant foregrounds have been removed from the recovered noisy Q map. We also plot the Planck foreground-cleaned Q map, from using the Commander method, in panel (a) of Figure \ref{figure3}, which is quite similar to the recovered noisy Q map that we obtain using the CNN. The recovered noisy U map also displays a similar result as for the recovered Q map. We do not provide any quantitative comparison here, since we do not have a "true" map of the pure CMB signal for the real Planck observations.

Besides the comparison at the map level, in panel (b) of Figure \ref{figure3} we also plot the EE and BB power spectra obtained from the recovered noisy Q/U maps (red lines), which are quite consistent with the Planck best-fit power spectra (green lines), together with the ILC noise and the beam at 143 GHz. Based on the results of these comparisons, we can draw the conclusion that the foreground contaminations in the Planck observations have been cleanly removed at both the map level and the power spectrum level by using the CNN network.


\begin{figure}
	\includegraphics[width=1.0\hsize]{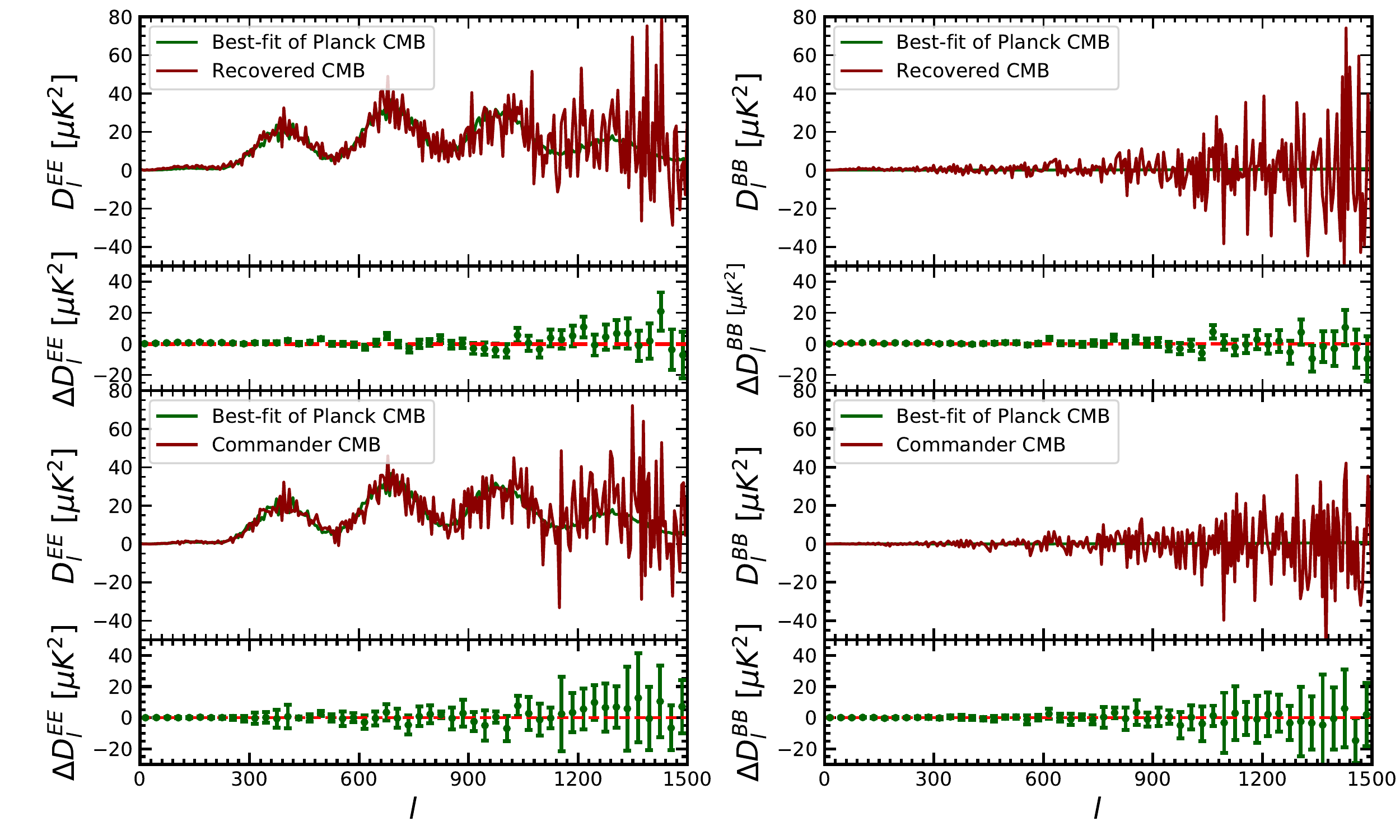}
	\vspace{-0.4cm}
	\caption{Upper panel: the cross-correlation power spectra of the recovered even-ring and odd-ring Planck CMB maps. Lower panel: the cross-correlation power spectra of the Commander foreground-cleaned maps. We also show the deviations of these cross-correlation power spectra from the Planck best-fit power spectra.}
	\label{figure4}
\end{figure}


Finally, we use the cross-correlation technique between two recovered HR maps to suppress the noise effects on the power spectra. In the two upper panels of Figure \ref{figure4}, we plot the recovered EE and BB power spectra, by cross-correlating two HR maps (red lines) with the Planck best-fit power spectra (green lines). Under each panel, we also plot the difference, $\Delta D_{\ell,{\rm CNN}} =D_{\ell,{\rm recover}} - D_{\ell, {\rm bestfit}}$, 
where the error bars represent the mean standard deviations over all the testing sets. We can see that the recovered power spectra are in good agreement with the Planck best-fit power spectra. We also calculate  $R^2$ and obtain $R^2=0.76\pm 0.39$ (68\% C.L.) for the recovered EE power spectrum, which also supports our conclusion. If we restrict our calculation at scales $\ell<1000$, the $R^2$ becomes better: $R^2=0.97\pm 0.05$ (68\% C.L.). The performance of recovery for EE power spectrum is quite good at $\ell<1000$. Here, we do not provide the $R^2$ value for the recovered BB power spectrum, since, due to the precision limitations of the Planck measurements, we cannot obtain reasonable constraints on the BB power spectrum.

In two lower panels of Figure \ref{figure4}, we also show the EE and BB power spectra obtained from the Commander foreground-cleaned map. We can see that these spectra are very similar to the recovered EE and BB power spectra obtained by using our CNN network, but with slightly smaller variance. We calculate  $R^2$ between the Commander EE and the best-fit EE power spectra: $R^2= 0.79$ ($\ell<1500$) and $R^2=0.97$ ($\ell<1000$), which are consistent with the values that we obtain for the recovered EE power spectrum.

\section{Conclusions}

In this paper, we present a machine-learning CMB foreground-cleaning technique, which involves two steps: (1) using the CNN network to remove the foreground contamination from the raw CMB map; and (2) using a cross-correlation technique to suppress the instrumental noise effects on the power spectra obtained from the recovered noisy CMB map.

We apply this pipeline to the CMB-S4 experiment. At the map level, CMBFSCNN can cleanly remove the polarized foreground components from both the Q and U maps. The MAD values between the recovered and target noisy maps are $0.018 \pm 0.023\ \mu$K for Q map recovery and $0.021 \pm 0.028\ \mu$K for U map recovery.  After dividing the data into two HS maps, we cross-correlate them, to suppress the noise effects at the power spectrum level. The calculations of $R^2$ for the recovered EE and BB power spectra prove that the CMB signal obtained by using our method is reliable and almost identical to the input fiducial CMB information. We also apply this method to Planck observations, to test its performance. The results show that  \texttt{CMBFSCNN} is capable of recovering precise CMB polarization signals from observational data sets, and that it can provide similar power spectra to those from the Commander foreground-cleaned maps, demonstrating that our method has the potential for CMB component separation and B-mode detection. 

There are still some limitations to the use of our method. The number of input frequency bands has a slight impact on the performance of our method. The performance of the recovered CMB map will be reduced if the network input is a map that has only been observed in one frequency band. In addition, the noise on an observed map with a higher pixel resolution ($N_{\rm side}=1024$ or $N_{\rm side}=2048$) will have a higher amplitude than the map used in this work ($N_{\rm side}=512$), which will slightly reduce the performance of our method when the network handles such high-resolution maps.

We will investigate the capability of our method in relation to other aspects, such as reconstructions of foregrounds and CMB delensing, as well as the component separation of future radio surveys, in future works.

\section*{Acknowledgements}

J.-Q.X. is supported by the National Science Foundation of China, under grant Nos. U1931202 and 12021003, and by the National Key R\&D Program of China, Nos. 2017YFA0402600 and 2020YFC2201603. Some of the results in this paper have been derived using the HEALPix (page: \url{https://healpix.sourceforge.io/}).

\bibliography{bibliography_recovered_cmb.bib}

\providecommand{\noopsort}[1]{}\providecommand{\singleletter}[1]{#1}%
\begin{thebibliography}{}
\expandafter\ifx\csname natexlab\endcsname\relax\def\natexlab#1{#1}\fi
\providecommand{\url}[1]{\href{#1}{#1}}
\providecommand{\dodoi}[1]{doi:~\href{http://doi.org/#1}{\nolinkurl{#1}}}
\providecommand{\doeprint}[1]{\href{http://ascl.net/#1}{\nolinkurl{http://ascl.net/#1}}}
\providecommand{\doarXiv}[1]{\href{https://arxiv.org/abs/#1}{\nolinkurl{https://arxiv.org/abs/#1}}}

\bibitem[{{Abazajian} {et~al.}(2019){Abazajian}, {Addison}, {Adshead}, {Ahmed},
  {Allen}, {et~al.}}]{Abazajian:2019}
{Abazajian}, K., {Addison}, G., {Adshead}, P., {et~al.} 2019, arXiv e-prints,
  arXiv:1907.04473.
\newblock \doarXiv{1907.04473}

\bibitem[{Ade {et~al.}(2019)Ade, Aguirre, Ahmed, Aiola, , {et~al.}}]{Ade2019}
Ade, P., Aguirre, J., Ahmed, Z., {et~al.} 2019, Journal of Cosmology and
  Astroparticle Physics, 2019, 056, \dodoi{10.1088/1475-7516/2019/02/056}

\bibitem[{{Alonso} {et~al.}(2019){Alonso}, {Sanchez}, {Slosar}, \& {LSST Dark
  Energy Science Collaboration}}]{Alonso2019}
{Alonso}, D., {Sanchez}, J., {Slosar}, A., \& {LSST Dark Energy Science
  Collaboration}. 2019, \mnras, 484, 4127, \dodoi{10.1093/mnras/stz093}

\bibitem[{Armitage-Caplan {et~al.}(2012)Armitage-Caplan, Dunkley, Eriksen,
  {et~al.}}]{Armitage-Caplan:2012}
Armitage-Caplan, C., Dunkley, J., Eriksen, H.~K., {et~al.} 2012, \mnras, 424,
  1914, \dodoi{10.1111/j.1365-2966.2012.21314.x}

\bibitem[{{Casas} {et~al.}(2022){Casas}, {Bonavera}, {Gonz{\'a}lez-Nuevo},
  {Baccigalupi}, {Cueli}, {Crespo}, {Goitia}, {Santos}, {S{\'a}nchez}, \& {de
  Cos}}]{Casas:2022}
{Casas}, J.~M., {Bonavera}, L., {Gonz{\'a}lez-Nuevo}, J., {et~al.} 2022, arXiv
  e-prints, arXiv:2205.05623.
\newblock \doarXiv{2205.05623}

\bibitem[{{Dumoulin} \& {Visin}(2016)}]{Dumoulin:2016}
{Dumoulin}, V., \& {Visin}, F. 2016, arXiv e-prints, arXiv:1603.07285.
\newblock \doarXiv{1603.07285}

\bibitem[{Eriksen {et~al.}(2008)Eriksen, Jewell, Dickinson,
  {et~al.}}]{Eriksen:2008}
Eriksen, H.~K., Jewell, J.~B., Dickinson, C., {et~al.} 2008, ApJ, 676, 10,
  \dodoi{10.1086/525277}

\bibitem[{{Errard} {et~al.}(2016){Errard}, {Feeney}, {Peiris},
  {et~al.}}]{Errard:2016}
{Errard}, J., {Feeney}, S.~M., {Peiris}, H.~V., {et~al.} 2016, JCAP, 2016, 052,
  \dodoi{10.1088/1475-7516/2016/03/052}

\bibitem[{{Farsian} {et~al.}(2020){Farsian}, {Krachmalnicoff}, \&
  {Baccigalupi}}]{Farsian:2020}
{Farsian}, F., {Krachmalnicoff}, N., \& {Baccigalupi}, C. 2020, \jcap, 2020,
  017, \dodoi{10.1088/1475-7516/2020/07/017}

\bibitem[{Fern\'andez-Cobos {et~al.}(2016)Fern\'andez-Cobos, Marcos-Caballero,
  Vielva, Mart\'\i{}nez-Gonz\'alez, \& Barreiro}]{Fernandez-Cobos:2016iud}
Fern\'andez-Cobos, R., Marcos-Caballero, A., Vielva, P.,
  Mart\'\i{}nez-Gonz\'alez, E., \& Barreiro, R.~B. 2016, Mon. Not. Roy. Astron.
  Soc., 459, 441, \dodoi{10.1093/mnras/stw670}

\bibitem[{{Finkbeiner} {et~al.}(1999){Finkbeiner}, {Davis}, \&
  {Schlegel}}]{Finkbeiner:1999}
{Finkbeiner}, D.~P., {Davis}, M., \& {Schlegel}, D.~J. 1999, ApJ, 524, 867,
  \dodoi{10.1086/307852}

\bibitem[{Fluke \& Jacobs(2020)}]{Fluke:2020}
Fluke, C.~J., \& Jacobs, C. 2020, WIREs Data Min Knowl Discovery, 10, e1349,
  \dodoi{10.1002/widm.1349}

\bibitem[{{G{\'o}rski} {et~al.}(2005){G{\'o}rski}, {Hivon}, {Banday},
  {et~al.}}]{Gorski2005}
{G{\'o}rski}, K.~M., {Hivon}, E., {Banday}, A.~J., {et~al.} 2005, ApJ, 622,
  759, \dodoi{10.1086/427976}

\bibitem[{Hazumi {et~al.}(2019)Hazumi, Ade, Akiba, {et~al.}}]{Hazumi:2019}
Hazumi, M., Ade, P. A.~R., Akiba, Y., {et~al.} 2019, Journal of Low Temperature
  Physics, 194, 443, \dodoi{10.1007/s10909-019-02150-5}

\bibitem[{Hensley \& Bull(2018)}]{Hensley:2018}
Hensley, B.~S., \& Bull, P. 2018, ApJ, 853, 127,
  \dodoi{10.3847/1538-4357/aaa489}

\bibitem[{Kamionkowski \& Kovetz(2016)}]{Kamio:2016}
Kamionkowski, M., \& Kovetz, E.~D. 2016, \araa, 54, 227,
  \dodoi{10.1146/annurev-astro-081915-023433}

\bibitem[{{Kim} {et~al.}(2009){Kim}, {Naselsky}, \& {Christensen}}]{Kim:2009}
{Kim}, J., {Naselsky}, P., \& {Christensen}, P.~R. 2009, \prd, 79, 023003,
  \dodoi{10.1103/PhysRevD.79.023003}

\bibitem[{{Kingma} \& {Ba}(2014)}]{Kingma:2014}
{Kingma}, D.~P., \& {Ba}, J. 2014, arXiv e-prints, arXiv:1412.6980.
\newblock \doarXiv{1412.6980}

\bibitem[{Kogut(2012)}]{Kogut:2012}
Kogut, A. 2012, ApJ, 753, 6, \dodoi{10.1088/0004-637x/753/2/110}

\bibitem[{Kogut {et~al.}(2007)Kogut, Dunkley, Bennett, {et~al.}}]{Kogut:2007}
Kogut, A., Dunkley, J., Bennett, C.~L., {et~al.} 2007, ApJ, 665, 355,
  \dodoi{10.1086/519754}

\bibitem[{Krachmalnicoff {et~al.}(2016)Krachmalnicoff, Baccigalupi, Aumont,
  {et~al.}}]{Krachmalnicoff:2016}
Krachmalnicoff, N., Baccigalupi, C., Aumont, J., {et~al.} 2016, \aap, 588, A65,
  \dodoi{10.1051/0004-6361/201527678}

\bibitem[{Krachmalnicoff {et~al.}(2018)Krachmalnicoff, Carretti, Baccigalupi,
  {et~al.}}]{Krachmalnicoff:2018}
Krachmalnicoff, N., Carretti, E., Baccigalupi, C., {et~al.} 2018, \aap, 618,
  18, \dodoi{10.1051/0004-6361/201832768}

\bibitem[{{Krachmalnicoff} {et~al.}(2022){Krachmalnicoff}, {Matsumura}, {de la
  Hoz}, {et~al.}}]{Krachmalnicoff2022}
{Krachmalnicoff}, N., {Matsumura}, T., {de la Hoz}, E., {et~al.} 2022, JCAP,
  2022, 039, \dodoi{10.1088/1475-7516/2022/01/039}

\bibitem[{{Krachmalnicoff} \& {Puglisi}(2021)}]{Krachmalnicoff:2021}
{Krachmalnicoff}, N., \& {Puglisi}, G. 2021, \apj, 911, 42,
  \dodoi{10.3847/1538-4357/abe71c}

\bibitem[{{Li} {et~al.}(2017){Li}, {Li}, {Liu}, {Li}, {Cai}, {Li}, {Zhao},
  {Liu}, {Li}, {Xu}, {Wu}, {Zhang}, {Fan}, {Yao}, {Kuo}, {Lu}, \&
  {Zhang}}]{Li2017}
{Li}, H., {Li}, S.-Y., {Liu}, Y., {et~al.} 2017, arXiv e-prints,
  arXiv:1710.03047.
\newblock \doarXiv{1710.03047}

\bibitem[{Mehta {et~al.}(2019)Mehta, Bukov, Wang, {et~al.}}]{Mehta:2019}
Mehta, P., Bukov, M., Wang, C.~H., {et~al.} 2019, \physrep, 810, 1,
  \dodoi{10.1016/j.physrep.2019.03.001}

\bibitem[{{Nah} {et~al.}(2016){Nah}, {Kim}, \& {Lee}}]{Nah2016}
{Nah}, S., {Kim}, T.~H., \& {Lee}, K.~M. 2016, arXiv e-prints.
\newblock \doarXiv{1612.02177}

\bibitem[{Petroff {et~al.}(2020)Petroff, Addison, Bennett,
  {et~al.}}]{Petroff:2020}
Petroff, M.~A., Addison, G.~E., Bennett, C.~L., {et~al.} 2020, ApJ, 903, 104,
  \dodoi{10.3847/1538-4357/abb9a7}

\bibitem[{{Planck Collaboration} {et~al.}(2016){Planck Collaboration}, {Adam},
  {Ade}, {et~al.}}]{PlanckCollaboration:2016}
{Planck Collaboration}, {Adam}, R., {Ade}, P.~A.~R., {et~al.} 2016, \aap, 594,
  A1, \dodoi{10.1051/0004-6361/201527101}

\bibitem[{{Planck Collaboration} {et~al.}(2014b){Planck Collaboration}, {Ade},
  {Aghanim}, {et~al.}}]{PlanckCollaboration:2014b}
{Planck Collaboration}, {Ade}, P.~A.~R., {Aghanim}, N., {et~al.} 2014b, \aap,
  571, A6, \dodoi{10.1051/0004-6361/201321570}

\bibitem[{{Planck Collaboration} {et~al.}(2014c){Planck Collaboration}, {Ade},
  {Aghanim}, {et~al.}}]{PlanckCollaboration:2014c}
---. 2014c, \aap, 571, A12, \dodoi{10.1051/0004-6361/201321580}

\bibitem[{{Planck Collaboration} {et~al.}(2014a){Planck Collaboration},
  {Aghanim}, {Armitage-Caplan}, {et~al.}}]{PlanckCollaboration:2014}
{Planck Collaboration}, {Aghanim}, N., {Armitage-Caplan}, C., {et~al.} 2014a,
  \aap, 571, A2, \dodoi{10.1051/0004-6361/201321550}

\bibitem[{Remazeilles {et~al.}(2016)Remazeilles, Dickinson, Eriksen,
  {et~al.}}]{Remazeilles:2016}
Remazeilles, M., Dickinson, C., Eriksen, H. K.~K., {et~al.} 2016, \mnras, 458,
  2032, \dodoi{10.1093/mnras/stw441}

\bibitem[{{Ronneberger} {et~al.}(2015){Ronneberger}, {Fischer}, \&
  {Brox}}]{Ronneberger:2015}
{Ronneberger}, O., {Fischer}, P., \& {Brox}, T. 2015, arXiv e-prints.
\newblock \doarXiv{1505.04597}

\bibitem[{Stompor {et~al.}(2016)Stompor, Errard, \& Poletti}]{Stompor:2016}
Stompor, R., Errard, J., \& Poletti, D. 2016, \prd, 94, 083526.
\newblock \url{https://dx.doi.org/10.1103/physrevd.94.083526}

\bibitem[{Sudevan {et~al.}(2017)Sudevan, Aluri, Yadav, {et~al.}}]{Sudevan:2017}
Sudevan, V., Aluri, P.~K., Yadav, S.~K., {et~al.} 2017, ApJ, 842, 62,
  \dodoi{10.3847/1538-4357/aa7334}

\bibitem[{Suzuki {et~al.}(2018)Suzuki, Ade, Akiba, {et~al.}}]{Suzuki:2018}
Suzuki, A., Ade, P. A.~R., Akiba, Y., {et~al.} 2018, Journal of Low Temperature
  Physics, 193, 1048, \dodoi{10.1007/s10909-018-1947-7}

\bibitem[{Tegmark {et~al.}(2004)Tegmark, Strauss, Blanton,
  {et~al.}}]{Tegmark:2004}
Tegmark, M., Strauss, M.~A., Blanton, M.~R., {et~al.} 2004, \prd, 69, 103501.
\newblock \url{https://dx.doi.org/10.1103/physrevd.69.103501}

\bibitem[{{Thorne} {et~al.}(2017){Thorne}, {Dunkley}, {Alonso},
  {et~al.}}]{Thorne:2017}
{Thorne}, B., {Dunkley}, J., {Alonso}, D., {et~al.} 2017, \mnras, 469, 2821,
  \dodoi{10.1093/mnras/stx949}

\bibitem[{Tian {et~al.}(2020)Tian, Xu, \& Zuo}]{Tian:2020}
Tian, C., Xu, Y., \& Zuo, W. 2020, Neural Networks, 121, 461,
  \dodoi{https://doi.org/10.1016/j.neunet.2019.08.022}

\bibitem[{Wang {et~al.}(2020)Wang, Li, \& Xia}]{Wang:2020}
Wang, G.~J., Li, S.~Y., \& Xia, J.~Q. 2020, ApJS, 249, 17,
  \dodoi{10.3847/1538-4365/aba190}

\bibitem[{{Wang} {et~al.}(2022){Wang}, {Shi}, {Yan}, {Xia},
  {et~al.}}]{Wang:2022}
{Wang}, G.-J., {Shi}, H.-L., {Yan}, Y.-P., {Xia}, J.-Q., {et~al.} 2022, arXiv
  e-prints.
\newblock \doarXiv{2204.01820}

\bibitem[{{Waqas Zamir} {et~al.}(2021){Waqas Zamir}, {Arora}, {Khan}, ,
  {et~al.}}]{Waqas2021}
{Waqas Zamir}, S., {Arora}, A., {Khan}, S., , {et~al.} 2021, arXiv e-prints.
\newblock \doarXiv{2102.02808}

\bibitem[{{Yu} \& {Koltun}(2015)}]{Yu:2015}
{Yu}, F., \& {Koltun}, V. 2015, arXiv e-prints, arXiv:1511.07122.
\newblock \doarXiv{1511.07122}

\bibitem[{Zacchei {et~al.}(2011)Zacchei, Maino, Baccigalupi,
  {et~al.}}]{Zacchei:2011}
Zacchei, A., Maino, D., Baccigalupi, C., {et~al.} 2011, \aap, 536, A5,
  \dodoi{10.1051/0004-6361/201116484}

\bibitem[{Zhang {et~al.}(2021)Zhang, Liu, Li, Wu, Li, \& Li}]{Zhang:2021odl}
Zhang, Z., Liu, Y., Li, S.-Y., {et~al.} 2021.
\newblock \doarXiv{2109.12619}

\bibitem[{Zonca {et~al.}(2019)Zonca, Singer, Lenz, Reinecke, Rosset, Hivon, \&
  Gorski}]{Zonca2019}
Zonca, A., Singer, L., Lenz, D., {et~al.} 2019, Journal of Open Source
  Software, 4, 1298, \dodoi{10.21105/joss.01298}

\end{thebibliography}
\bibliographystyle{aasjournal}

\end{document}